\documentclass[journal]{IEEEtran}
\usepackage{graphicx}
\usepackage{cite}
\usepackage{balance}
\usepackage{caption}
\usepackage{amssymb}
\usepackage[colorlinks]{hyperref}

\ifCLASSINFOpdf
\else
\fi
\begin{document}
%
\title{\textcolor{red}{Blockchain Technology with Applications to Distributed Control and Cooperative Robotics: A Survey}}
\author{Ameer Tamoor Khan, Xinwei Cao, and Shuai Li, \IEEEmembership{Senior Member,~IEEE}
\thanks{This work is supported by Hong Kong Research Grants Council Early Career Scheme (with number 25214015), by Department General Research Fund of Hong Kong Polytechnic University (with number G.61.37.UA7L), and also by PolyU Central Research Grant (with number G-YBMU), by National Natural Science Foundation of China (with number 61503413))}	
\thanks{ A. Khan, S. Li with the Department of Computing, The Hong Kong
	Polytechnic University, Hung Hom, Kowloon, Hong Kong, China, and Prof. Xinwei
	Cao is from Shanghai University, China (Corresponding author, e-mail:
	shuaili@polyu.edu.hk)
}
}

\maketitle
\markboth{}%
{Shell \MakeLowercase{\textit{et al.}}: Bare Demo of IEEEtran.cls
	for Journals}
\maketitle


\maketitle

\begin{abstract}
  As a disruptive technology, blockchain, particularly its original form of bitcoin as a type of digital currency, has attracted great attentions. The innovative distributed decision making and security mechanism lay the technical foundation for its success, making us consider to penetrate the power of blockchain technology to distributed control and cooperative robotics, in which the distributed and secure mechanism is also highly demanded. Actually, security and distributed communication have long been unsolved problems in the field of distributed control and cooperative robotics. It has been reported on the network failure and intruder attacks of distributed control and multi-robotic systems. Blockchain technology provides promise to remedy this situation thoroughly. This work is intended to create a global picture of blockchain technology on its working principle and key elements in the language of control and robotics, to provide a shortcut for beginners to step into this research field.
  
\end{abstract}


%
\IEEEpeerreviewmaketitle

\section{Introduction}\label{s1}
The blockchain is a decentralized and peer-to-peer network applicable to eliminate third-party interference or involvement in business, and data sharing activities. It is originated from bitcoin and has been much beyond that by penetrating to various transactions, including money transfer, data sharing, secret transmission, thanks to its inherent nature of low latency and immunity against attacks  \cite{1}.\par
It is a disruptive technology that overthrows the conventional business model with a low-cost solution. It allows its peers to formulate or legislate their terms of business, and once conditions are formulated and a company is in a run, it will provide the most secure environment for trade, commerce, and dealings. For example, a matter like stock exchange can be done within a duration of microseconds with involving human intervention. However, in real-world, this process will take many intermediary steps because parties do not have access to each other's ledgers to verify the validity of their transactions. Hail to the blockchain, who made this process so easy because of its decentralization \cite{2} and secure peer-to-peer connection. \par
The blockchain was first introduced in 1991 by S. Haber and W. Stornetta but didn't the popularity back then. This concept was first conceptualized by the Satoshi Nakamoto in 2008, by introducing cryptocurrency bitcoin \cite{3}. Soon this concept got momentum and was spread widely among the business and commerce community \cite{5}.
Those unique features of blockchain are also highly demanded by distributed control systems, which are usually complex systems for plant control with sensors and actuators geographically different locations. Conventional DCS is evolved from centralized power with a vulnerable central control station for the monitoring of the whole system and suffers from potential risks, e.g., strict security and privacy issue, with a threat leading to the hijack by intruders. If a single unit fails to perform it will not affect the whole system, as compared to a centralized system where everything is connected to a common node, failure in one unit will bring down the entire system. It still faces some potential security and privacy threats like; interception, interruption, modification, fabrication, encryption, authentication and auditing. The fully distributed nature and security guarantee of blockchain endow it potentials in DCS applications. Its decentralization is the critical element for the security check.\par

\begin{figure}
	\centering
	\includegraphics[width=1\linewidth]{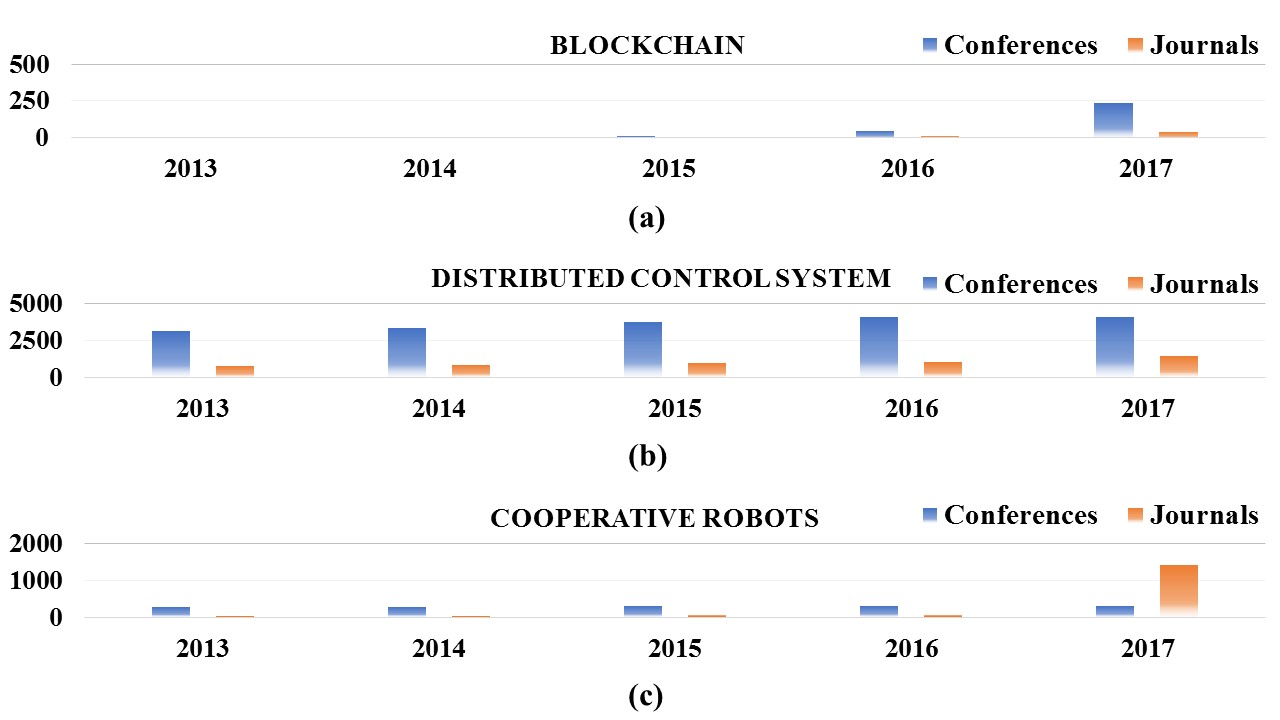}
	\caption{Figs. (a), (b), and (c) show the rising trend in research of blockchain, DCS and cooperative robots. Fig. (a) shows that research increases up-to 238\% in blockchain from 2013 to 2017. Fig. (b) shows that the research activities in DCS has been quite a lot since 2013 and has been increasing steadily till 2017. Fig. (c) shows the research trend in cooperative robots, and it increased sharply in recent years. \vspace{-0.5cm}}
	\label{Fig.1}
\end{figure}

Multi-robots can handle complex tasks in a cooperative environment, despite the simple structure and limited capability of each robot. It consists of homogeneous robots with more or less same capabilities, fully autonomous to effectuate a task.  However, one restricting factor of cooperative robots is the security concern. An intruder robot can quickly get into an army of robots to create misconception between them if the coordination algorithm lacks a proper security mechanism. Similarly, decision making can also raise questions related to security matters, an army of swarm robots need an adequate platform to make decision unanimously without confusion, this problem can cause severe damage to the swarm robots when they are on a highly intelligent mission. \par
In this survey, we have discussed the disruptive technology of blockchain and how it makes the transaction of information security in a distrustful environment, without intervening third-party. Under the framework of the blockchain, we discuss distributed networks and investigate possible ways to make them distributed, raise concerns about security and other issues they have, cast lights on how blockchain facilitates them to overcome these challenges.and demonstrate potential applications of blockchain technology in DCS and cooperative robotics.
A lot of research has been done in these three fields; blockchain, Distributed control and cooperative robots. From Fig.\ref{Fig.1} it is quite obvious that all three have a rising graph for the last couple of years. Bitcoin Fig.\ref{Fig.1}(a) is leading this race, from 2013 to 2017 the percentage of a research paper has reached up-to 238\%. This set of data reflects the increasing attention and popularity of blockchain technology and its impact in a research community. Same is the trend with DCS and cooperative robots as shown in Fig.\ref{Fig.1}(b) and Fig.\ref{Fig.1}(c) respectively.  
\section{Overview of Blockchain Technology}\label{s2}
The blockchain is a disruptive technology that is based on a decentralized system, meaning that it is a free-form third-party monopoly. It has a distributed database that is duplicated among all peers in the network. It creates a trust-worthy, meddle-free environment among mutually untrusted peers. It serves as a business transaction ledger, a system for the sharing of data, assets and valuable records, by employing “proof of work” or another consensus mechanism to create a trustful, accountable, transparent and self-accountable environment instead of relying on some third-party mediator or actor. Since it replicates the data among all peers, it allows users to verify and audit the transactions inexpensively \cite{1,2,3}.\par
Bitcoin was the first peer-to-peer decentralized cryptocurrency that uses blockchain to record all transactional data. Satoshi Nakamoto first introduced it in 2008. After captivating the attention of the business community and lawmakers, bitcoin became increasingly popular and was accepted as a cheap, safe, open and reliable method of moving assets on the Internet, free from third-party interference. Around 4 million peers are connected to the network, and about 125,000 transactions are carried all around the world under the realm of bitcoin. Blockchain has gained a lot of researchers attention these years. As seen in Fig.\ref{Fig.1}, Research has been done in this field tremendously. It is also remarkable that from 2013 to 2017, the percentage of conference paper increases up-to 238\%. This set of data reflects the increasing attention and popularity of blockchain technology, its impact, and recognition in a research community.

\subsection{Key Elements of Blockchain}
In this section, we will understand the key elements that are present in a blockchain, there importance and how they make blockchain reliable for the storage and transfer of data.
\subsubsection{Block of a Blockchain}
Blockchain consists of blocks, and each block contains the head and the body. The body accommodates the data. In the case of bitcoin the collection includes the transactional data, and the head consists of metadata, such as a hash of the block, a hash of the previous block, proof of work and timestamp. Blocks are time bounded in a collection of data which means that after a specific time these blocks will become a permanent part of the blockchain and their data is not retractable or alterable. For example, in the case of Factor and bitcoin, each block is added to the chain after every 10 minutes. In the case of Ethereum, this time is 15 seconds.

\subsubsection{Chain in Blockchain}
Blockchain consists of a chain that connects blocks, thinking of a chain as an electrical glue that holds all the blocks and creates trust among them, “hash of the previous block” in each block connects it with the preceding block, which traces back to the very first block, the genesis block. Hash is a digital fingerprint for any block, it identifies the block and all its contents, and it’s always unique. If one succeeded in tempering the block, this will change the hash of that block too, meaning that all subsequent blocks become invalid because they no longer store the valid hash of the previous block and thus the chain breaks.

\subsubsection{Network of Blockchain}
\begin{figure}
	\centering
	\includegraphics[width=1\linewidth]{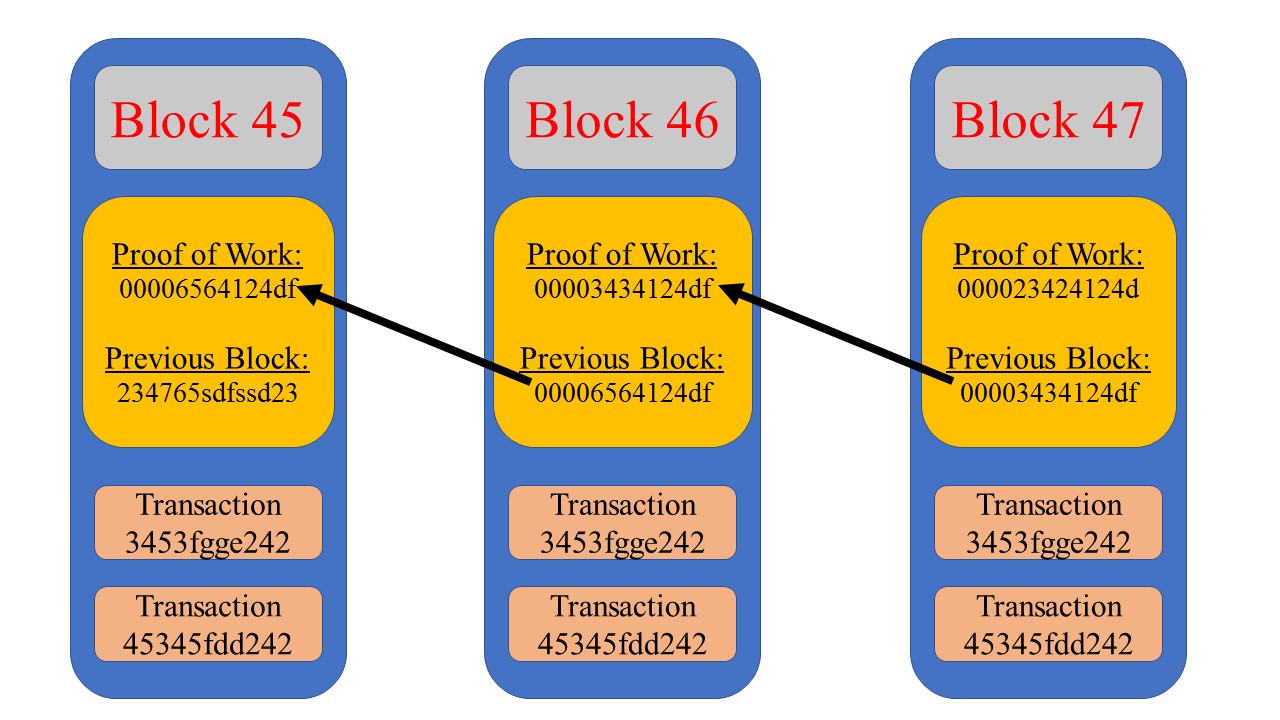}
	\caption{This a portion of a blockchain, showing how blocks are chained together through their ``proof of work". If someone tries to pollute block by altering its data, the hash of the block will be changed, breaking its connection with the previous block, i-e., breaking the chain, making the intruders move nullified and void. \vspace{-0.5cm}}
	\label{fig1.}	
\end{figure}

The network is the last and one important part of any blockchain. It is a collection, or a group of computers commonly known as a node. Node is simply a personal computer or any other system operating on the network. Defining nodes on the internet means anything with IP address. When any transaction occurs, or enter into the system, it is first verified by all the nodes, and once the consensus is made among most of the nodes, that transaction becomes the permanent part of the blockchain. The nodes get the labor for running the protocol with the portion of the cryptocurrency because it consumes their electrical-energy. To make the blockchain more secure, the network needs to be bigger and more distributed. The security of this system is related to decentralization, and it becomes more secure if it is scaled up. As the network grows, it becomes extremely hard for the hacker or attackers to overthrow the network. For example, a bitcoin network with 5000 nodes distributed globally is usually considered as an extremely secure one. The basic architecture of blockchain is shown in Fig. \ref{fig1.}. As observed in this figure, every block of the blockchain is connected with the previous block through ``proof of work", if anyone tempers any block the hash of that block will change too, disconnecting it from the remaining chain, making his wrong move void. In short, blockchain is powerful mainly because of two things; it’s chained blocks that make the data immutable and its decentralized network.

\subsection{How Blockchain Works}
A blockchain is a scalable, highly decentralized peer-to-peer system. It works on the mechanism of consensus rather than the trust. The decentralized peer-to-peer system eliminates the monopoly of the third party; it prevents the single user or the group of users from meddling or taking-over the system for their gain. All the participants are of equal status abiding by the same protocol. They can be a group of random people, governments or private organizations or the combination of them. To tie the knot, five basic principles underlying the working of the blockchain.
\begin{enumerate}
	\item  \textbf{Distributed Database: }The database is in access of every participant because of its peer-to-peer nature. No single party controls the data so that every individual can verify the record of its transactions and assets without third-party intervention or involvement.
	\item \textbf{Peer-to-Peer Transmission: }All the communication occurs between the participants or the parties directly without the intervention of the third party. When a transaction occurs, each node stores that transaction and forward it to other nodes.
	\item \textbf{Transparency: }When a transaction occurs all the information related to it, including the value associated with it, is sent to all the participants of that network. Each node has a unique 30-character alphanumeric address, which is its identity. It’s up to the user whether to remain anonymous or to provide its proof of identity to other users while making the transaction.
	\item \textbf{Ir-reversibility of Record: }It is impossible to alter the record if once the transaction is entered in the database and accounts are updated, it's impossible to retract the database. If someone tries to change the data and succeed in doing so, then the hash of that block will be changed, breaking its chain with the subsequent blocks because no subsequent block contains the valid hash of the preceding block. Various advance computational algorithms are employed to make sure that data remains permanent, chronologically ordered in a database, and remains available to all the peers of the network.
	\item \textbf{Computational Logic: }The ledger of the blockchain is in digital domain which means that it works under the logic. This allows the peers to set up their own rules and regulation to carry the transactions \cite{4,5,6}.    
\end{enumerate}

\subsection{Advantages of Blockchain}
In this section, we will discuss the advantages associated with this disruptive technology. Although there are many useful features of the blockchain here, we will focus on two of those, decentralization and security.

\begin{enumerate}
	\item \textbf{Decentralized Network: }Decentralization is the core feature of this technology, it liberates its users from the third-party interference and facilitate them with the benefits like trustlessness, full assess to data and immutability. Due to decentralization no intelligence agency, government, hackers or terrorist organization can take down the data because the replicated database is with all the peers of that network.    Proof of the work is one of the important consensus mechanism used to synchronize the millions of nodes relies on the complex mathematical problem that requires large computational power. This ensures that money remains safe and no duplicate transaction occurs \cite{8}.
	\item \textbf{Security and Privacy: }The name of Blockchain implies that it consists of the blocks connected in a chain. Each block consists of the hash of the previous block, so if someone alters the data of the block, this will change the hash of the block thus the chain will break. To temper the data, it is necessary that one changes the subsequent blocks as well so that chain remains intact. This is where the second firewall comes since the copy of ledger is available to all the peers, it means the alteration of one node is not enough. The attacker needs to take over at least 51\% of the nodes at the same time, this is known as a 51\% attack or inevitable attack. To avoid this attack, the networks to accommodate, network with fewer peers are vulnerable to this attack. If by any luck attacker passes this barrier then there comes the third defense, which is a ``proof of work", it requires the solution of a complex mathematical problem which needs both the computational, electrical power and time \cite{9}.
\end{enumerate}

\subsection{Application of Blockchain}
Blockchain has potential as well as a future. It’s been widely used for the digital currency or cryptocurrency; Bitcoin is an example of it. Land registry is another field where blockchain has made its place, countries like Georgia and Sweden is already working on it and India is also fighting land fraud with the help of blockchain. Swarm or cooperative robots face some problems which are tackle-able with the help of Blockchain, like decentralization of swarm robots, secure communication between them, security, distributed decision making, behavior differentiation, {\it etc.} There are other potential fields as well were one can employ the blockchain to make the system and its data decentralized, immutable, secure and open to its peers. 

\section{Distributed Control Systems}\label{s3}
In this section, we will explore the distributed control systems (DCS) and will learn about their architecture, how they are different from a centralized system, their applications, security threats and their remedy under the hood of a blockchain.

\subsection{What is Distributed Control}
\begin{figure}
	\centering
	\includegraphics[width=1\linewidth]{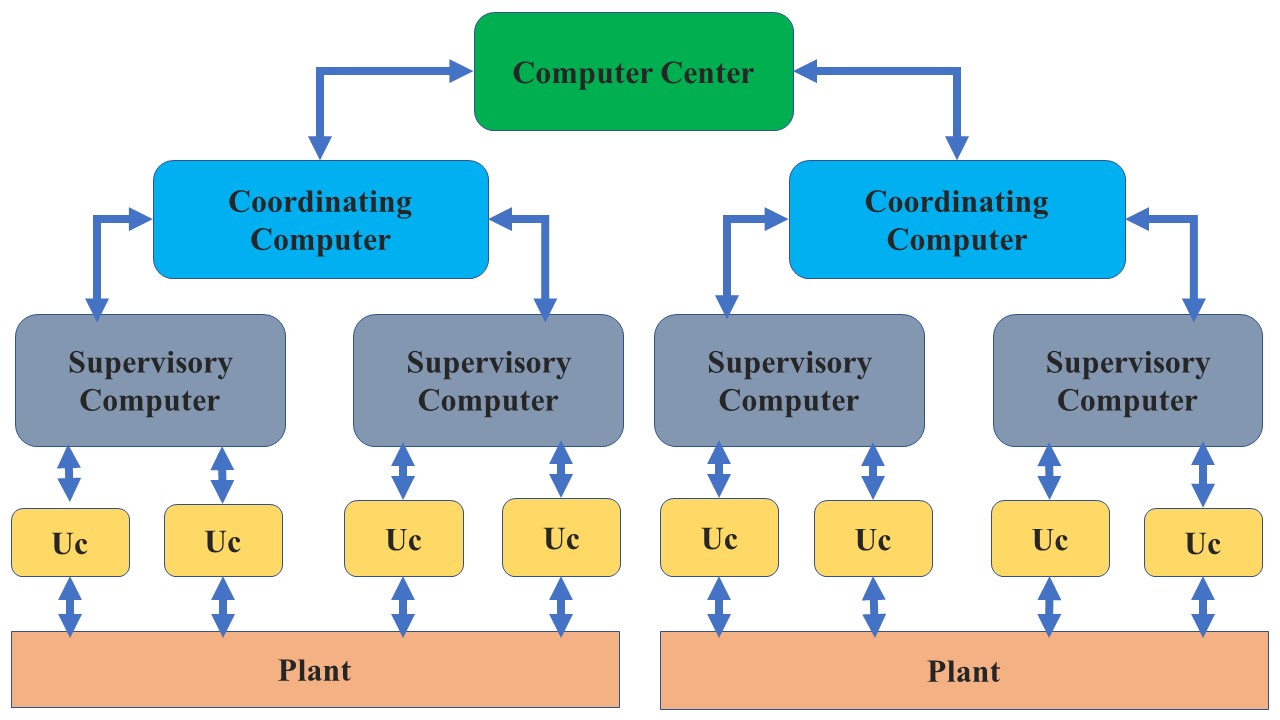}
	\caption{The figure shows, how DCS works in a hierarchal manner, sending info back and forth. It shows how the plants i-e sensors and actuators, since the environment and send the signals to micro-controllers, they further send the signals to the industrial units so that they can modify according to the need. Supervisory computers collect the information from the controllers and display it to the operator, Coordinating computers do not do anything instead, they only monitor the production units. Lastly, is the production scheduling.\vspace{-0.5cm}}
	\label{fig2.}
\end{figure}

A DCS is a fully autonomous and automated system that consists of control units distributed globally to control the complex, large and globally distributed applications, and industrial units. It is different from the centralized system, where a single central unit handles the control function, but in DCS every only group has its dedicated processor, these processors combined to form the control unit, and they are connected through a speed communication channel. In DCS, data accusation and the control tasks are performed via some microprocessors that are located near the control area. These controllers can communicate with each other as well as other controlling units \cite{10, 11, 12}. In DCS, controllers are connected to different field devices like actuators and sensors, they continuously receive data from them and send the data to other controllers in the hierarchy through a communication bus. Various communication channels are used for this purpose some of them are Profibus, HART, arc net, Modbus, {\it etc.} DCS is being employed in different walks of life which includes, agriculture, chemical plants. Petrochemical and refineries, nuclear power plants, water treatment plants, sewage treatment plants, food processing, automobile manufacturing, pharmaceutical manufacturing. The basic structure of the DCS is shown in Fig. \ref{fig2.}, that how it works in a hierarchal manner from top to bottom and vise versa, instead of a centralized system.

\subsection{Architecture of Distributed Control System}
The Architecture of the DCS has three main qualities.  The first one is the distribution of the control function into various sub-systems, which are connected through a high-speed communication channel. Data acquisition, data processing, control signals, reporting information are some of the main functionalities of these systems \cite{14}. 
The second quality is the provision of an autonomous system by integrating advanced control methods and algorithms, and the third attribute is the making of a system, which means that DCS contains different subsystems working under some control logic and command structure, that logic and structure unifies all the sub-systems \cite{15}. The Architecture of the DCS consists of four basic elements which are discussed below:
\begin{enumerate}
	\item \textbf{Engineering PC or Controller: }This is the supervisor controller where all the control logic and data are processed. It is connected with all the other controllers of the distributed system. A configuration of various devices and control algorithms are executed in this controller. The communication between the controllers and the Engineering PC can be done either through simplex or the redundant configuration.
	\item \textbf{Distributed or Local Control Unit: }This unit is located near the field devices, sensors, and actuators and is connected with them via cables and other communication channels. They receive control instructions and parameters from the engineering PC and directly controls the field devices. AC 700F and AC 800F are examples of local control instruments.
	\item \textbf{Operating Station or HMI: the }This purpose of this unit is to monitor the whole plant or system and display its performance on the monitor. Its other functionalities include data logging in the database of the plant, monitor the parameters of field devices \cite{17}.\\
	The Operating systems include PCs, some for the graphical display of different functionalities and some for the data log and some to monitor the parameter of the field devices.
	\item \textbf{Communication Media and Protocol: }Communication media includes the transmission cables, coaxial cables copper wires, fiber optics and even wireless communication. Different communication protocols can facilitate a different number of devices, for example, RS232 can connect two devices at a time whereas, Profibus can fcili126 devices for the same purpose.
	
\end{enumerate}

\subsection{Important Features of DCS}
\begin{figure}
	\centering
	\includegraphics[width=1\linewidth]{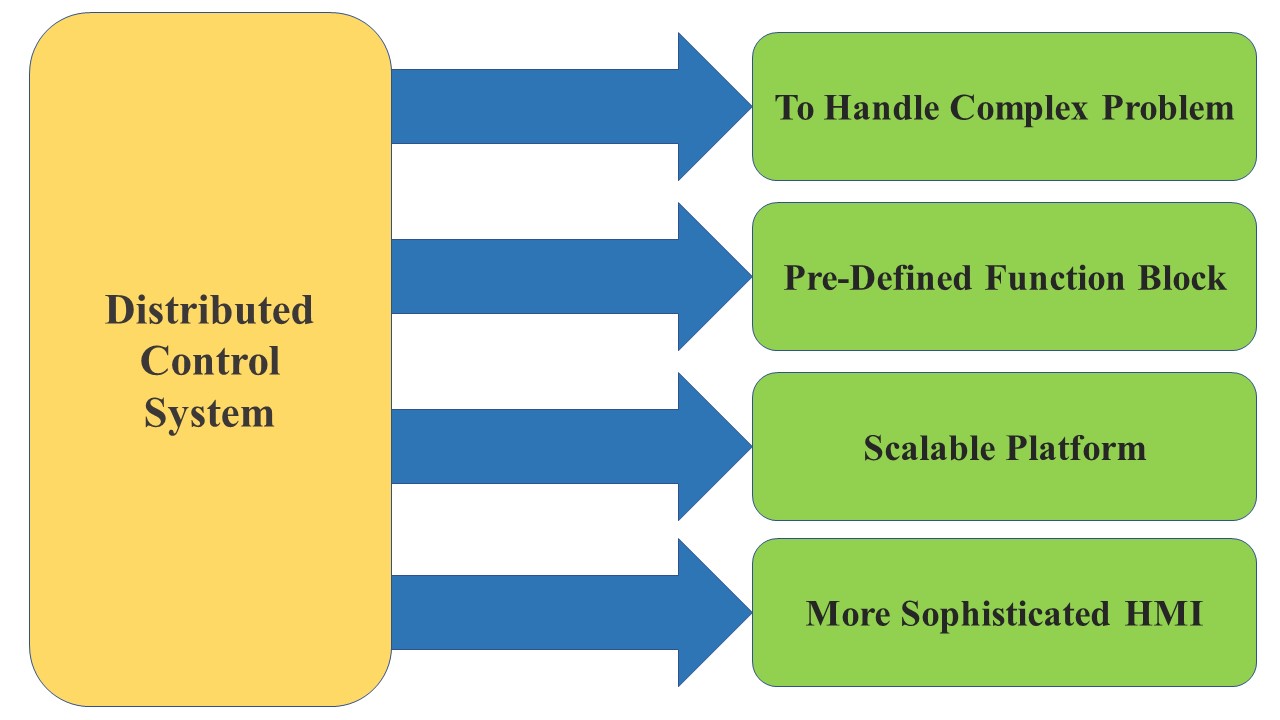}
	\caption{Some important features of DCS. Decentralized, and robustness are the core features that set them apart from centralized control systems. \vspace{-0.5cm}}
	\label{fig4.}
\end{figure}

In this section, we will explain some of the key features of the DCS which includes, handle complex processes, predefined function blocks, scalable platform, and More sophisticated HMI. The important features of DCS are shown in Fig. \ref{fig4.}, these features make them different from centralized systems.

\begin{enumerate}
	\item \textbf{To Handle Complex Processes: }In industries controllers like PLC are used to control and monitor the processes and parameters of different industrial plants at high-speed requirements.
	Hence these systems are preferred for the complex computations.
	\item \textbf{Pre-defined Function Blocks: }DCS offers pre-defined programs and algorithm which make its use more accessible for the complex problems and applications.
	It supports a large number of programming languages like function block, ladder, sequential, {\it etc.} so that a user can develop a custom program as per his requirement.
	\item \textbf{Scalable Platform: }DCS systems are scalable, it depends on the number of I/O’s connected to it. To accommodate more users, increase the input-output devices and the server to facilitate them.
	\item \textbf{More Sophisticated HMI: }DCS provides the HMI’s (Human Machine Interface) which allows the user to control or alter any process a per requirement.
	It displays the numerical and graphical performance of the system which helps the user to use that data and alter the conditions of the plant accordingly. 
\end{enumerate}

\subsection{Applications of DCS}
The area of its exercise and implementation is vast; it accommodates all the applications in the digital domain. Some of its applications are; drum level control, steam temperature control, furnace pressure control, combustion control, CBD tank level control, deaerator level, and pressure control, and PRDS control. The range of application also encircles airlines, banks, industries, manufacturers, retailers, and government organizations. Some other practical applications from our previous research work include; Distributed recurrent neural networks for cooperative control of manipulators, Distributed source seeking by cooperative robots.

\subsection{Security in Distributed Systems}
Security is a key feature that all systems should accommodate. In this section, we will discuss the security of the DCS.\par
In general, DCS faces these four types of security issues, interception, interruption, modification, and fabrication.
\begin{enumerate}
	\item \textbf{Interception: }An attempt to gain Unauthorized access to a system. An, e.g. attack on the server, illicit copying.
	\item \textbf{Interruption: }Unavailability of services and data may be because of someone intervention or tries to take over the server. E.g., deliberate corruption of a file, implantation of a virus, denial of service attacks.
	\item \textbf{Modification: }An illicit attempt to change the data such that it no longer remains original.
	\item \textbf{Fabrications: }It is an addition of unexcited data to the original data to make it lt looks original, e.g., addition of entries to the data, automatic reply to the messages, make them look original.\par
	To tackle these threats following security techniques are used, Encryption, Authentication, authorization, and auditing, but these techniques have their limitations.
	\item \textbf{Encryption: }It changes or transform the data from one domain to other, makes it difficult or even impossible for an attacker or hacker to understand.
	\item \textbf{Authentication: }It is used to authenticate that the user or peer connected to the system or server has the authority to access it.
	\item \textbf{Authorization: }It allows the rightful users to perform the specific tasks, and will block the way for intruders.
	\item \textbf{Auditing: }It is like any other audit system, used to maintain the check and balance of the system. It will keep track of how and who access the server.\par
	
\end{enumerate}

\subsection{Security threats remedy using Blockchain}
In this section, we will see how the disruptive technology of blockchain can provide security to the DCS systems. 
\begin{enumerate}
	\item Decentralization: As mentioned in \ref{s2}, decentralization is one of the key features of the blockchain, it liberates the system from the third-party interference. Security threats like interception and interruption are being tackled by employing blockchain for Distributes control system, because decentralization replicates the data to all the peers connected to the system, so in case someone tries to alter one copy of data, the remainder remains intact, and there is no concept of illegal copy of data, since blockchain is an open source.
	\item Blockchain to Tackle Modification, and Fabrication: In \ref{s2} we have discussed the key elements of Blockchain, which includes chain, means that it contains block connected with a chain. It is one of the firewalls provided by the blockchain to its database; each block contains the hash of the previous block which means that if someone tries to alter the data of any block then the hash of that block. Which means that the subsequent block no longer holds the valid hash of its preceding block. Thus, it breaks the chain and makes the attack unsuccessful. Back to DCS attack, if someone tries to modify the database, this firewall will tackle the problem and will keep the data safe and intact.
\end{enumerate}

\begin{figure}[tbh]
	\centering
	\includegraphics[width=1\linewidth]{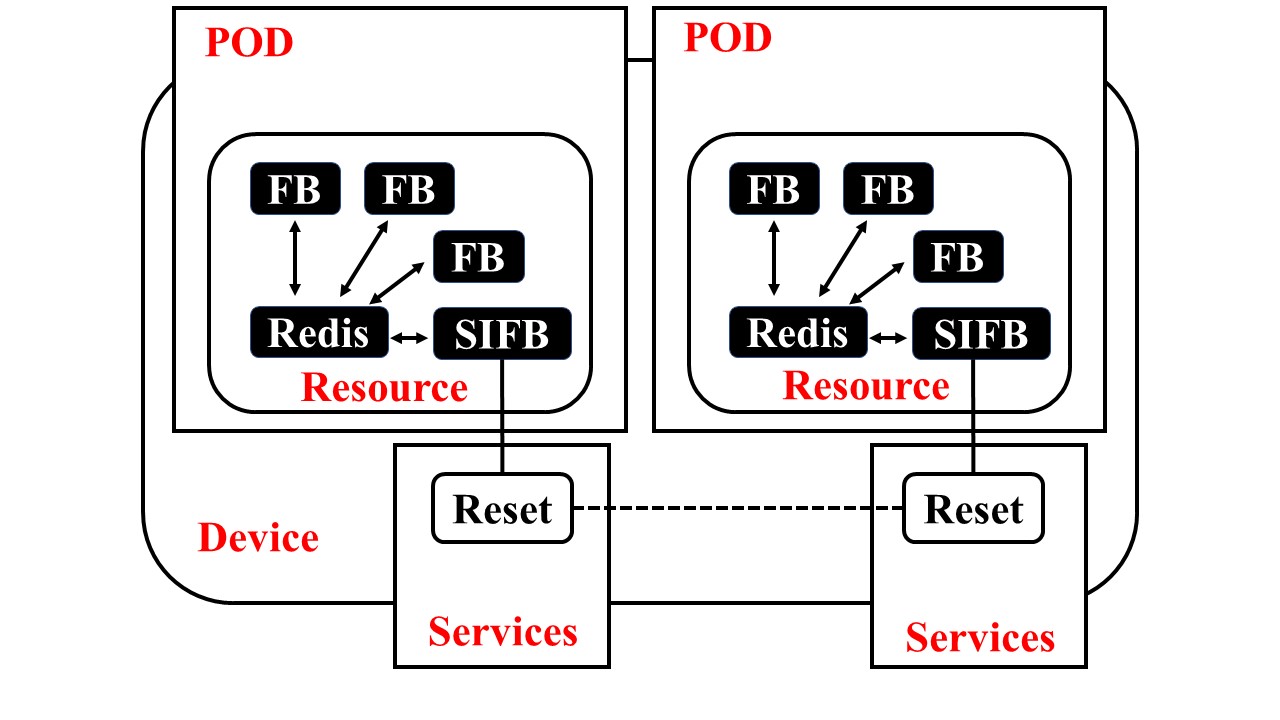}    
	\caption{\textcolor{red}{High-level architecture of IEC 61499 presented to implement PID control for dynamic system}}
	\label{exp1}
\end{figure}

\textcolor{red}{\subsection{Experimental results}
In this section, we will explain some experimental results to show how researchers have made use of blockchain in distributed control systems. We will explain two experiments; use of blockchain in edge computing \cite{76}, and blockchain for industrial system cyber-security \cite{77}.
\subsubsection{Blockchain in edge computing}
The work presented by Alexandru et al. cite{76} on edge computing; a novel method to provide computational power at the end of devices connected with the network used blockchain to give a hierarchical platform and distributed systems based on 
IEC 61499 standard. For blockchain, they implemented Hyper-ledger fabrics where functions are performed in small blocks or contracts on a supervisor level. \\
The case that they selected for integration of blockchain with the distributed control system is PID control for dynamic systems. The implementation of PID at a lower level is possible using three-tire edge computing, and on a higher level the algorithm or methods are responsible for the fine tuning of the controller. The provided system is shown in Fig.\ref{exp1}. 
}
\textcolor{red}{
The main aspect was that no small block within the system could assess the outside data such as APIs or other external services. To achieve that a consensus needs to be achieved between all the nodes of the blockchain.\\
To work around this problem the decentralized application of blockchain is used, which is used to push the data in blockchain so that the rest of the small blocks can consume it and make the data public. They evaluated the Hyperledger Fabric performance, based on the number of the made transactions per second (tps). They used the Google cloud resources i-e, \textit{n1-standard-4} and \textit{n1-standard-4}, for the benchmarking.\\
The results showed that the control distributed systems that are implemented using blockchain has a clear limitation to process the ample amount of data in real-time. Also, since the decisions made are based on the unanimity of the blockchain nodes, which ensures the secure medium for the transactional data. 
\subsubsection{Blockchain for industrial system cyber-security}
The work presented by Mao et al. \cite{77} implemented blockchain for the cyber-security of the industrial control system (ICS). As in blockchain, all the connected client's nodes share the copies of data. Similarly, they assume that the device node in ICS can be taken as blockchain node, every device nodes sends the data to the network, so they have taken advantage of this trait of the blockchain.\\
For the attack and defense experiment they implemented ICS blockchain network, let the devices share the data mutually and individually to each other and allows the blockchain to store the data.}
\section{Cooperative Robots}\label{s4}
In this section, we will discuss the cooperative or swarm robotic systems, their Applications, how they work, security issues and how we can employ blockchain to tackle these security issues.

\subsection{What Are the Cooperative Robots}
\begin{figure}
	\centering
	\includegraphics[width=1\linewidth]{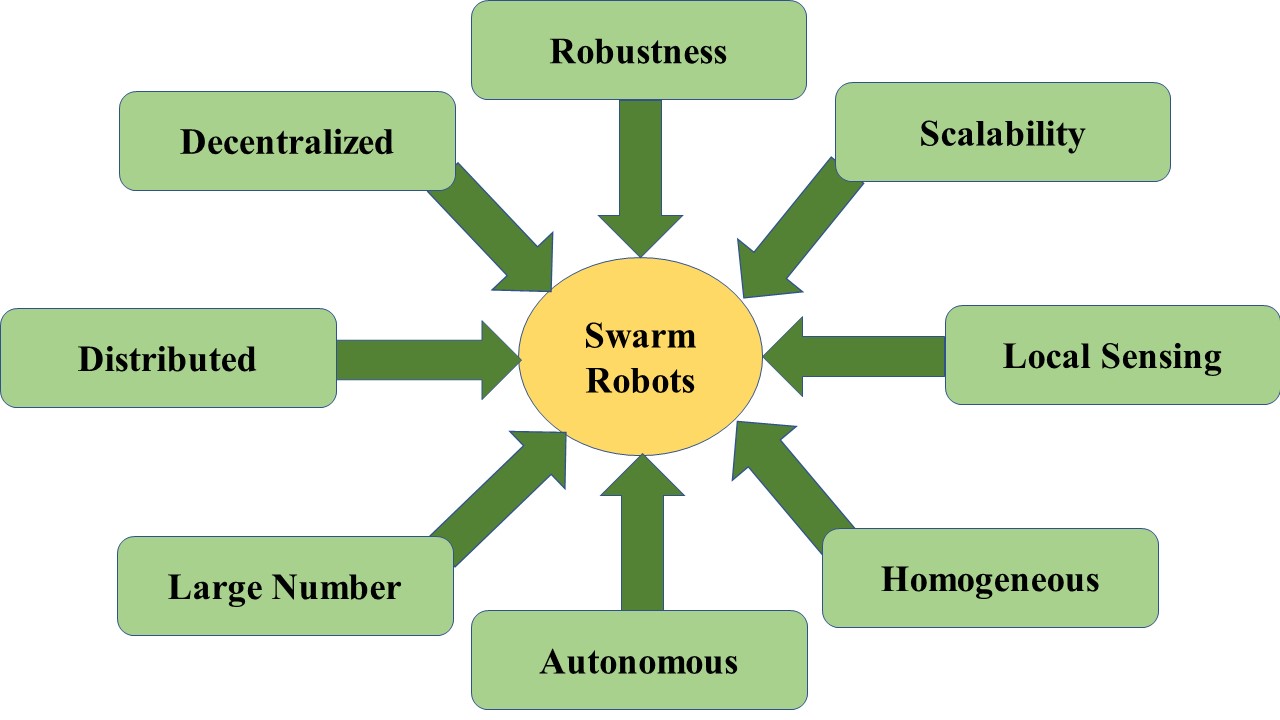}
	\caption{Characteristics of the cooperative or swarm robots. Distribution, decentralization and robustness are the main features that make them different form single robotic system. \vspace{-0.5cm}}
	\label{fig3.}
\end{figure}

Cooperative or swarm robots are the army of robots, trained to perform a complex task that is beyond the capacity of a single robot, like the automation industry, espionage activities, surgical purposes, {\it etc.}  \cite{18} Cooperative or swarm robots are the results of the inspiration from the insect society, where all the insects are identical, cooperatively perform tasks, in a decentralized way \cite{19}. The characteristic of the swarm robots is shown in Fig. \ref{fig3.}\par

The main characteristics and features of the cooperative robots are as follows:
\begin{enumerate}
	\item Cooperative robots should be fully autonomous, and able to sense the environment.
	\item They must consist of an army of swarm robots, operate under the intelligent control algorithm.
	\item The robots should be homogeneous, they can be a bit different, physically, and functionally but not much. 
	\item They should be incompetent and inefficient to execute any task, which means that they should work cooperatively to accomplish a task, to improve their performance and efficiency. 
	\item They have a private communication and sensing abilities, they have a distributed controlling mechanism, in which each robot will sense its task.
\end{enumerate}

\subsection{Single Robot Vs. Multiple Robots}
Here we will discuss the advantage and disadvantages of the cooperative robotic system over a single robotic system, the advantages of cooperative robots are stated below:

\begin{enumerate}
	\item \textbf{Improved Performance: }Large and complex tasks can be decomposed into sub-tasks, then by applying parallelism, we can divide the tasks among the robots. This will enhance their performance.
	\item \textbf{Enablement of Task: }Cooperative or swarm robots can accomplish, more complex and, difficult tasks that a single robot cannot.
	\item \textbf{Wider Sensing Range: }The sensing range of a swarm robotic system is wider than, a single robot.
	\item \textbf{Action Distributed: }In swarm robots, different robots can perform different tasks thus, their range of action is wider than, a single robotic system.
	\item \textbf{Enhanced Tolerance: }In a group, if a single robot faced a failure, it does not imply that a given task cannot be completed. The presence of the redundancy in the system will take care of it.
\end{enumerate}
Although there have been various advantages of cooperative robots, some disadvantages also exist: 
\begin{enumerate}
	\item \textbf{Interference: }Being in a group means that they can cause trouble or problems for each other, either through collisions or, through occlusions.
	\item \textbf{Uncertainty: }This system can work effectively if a robot knows about its fellow robot. If some uncertainty occurs than it can create a competitive and aggressive environment instead of cooperative.
	\item \textbf{Increase System Cost: }To make an army, it is necessary to put in more money, which increases the overall cost of the system. Usually, this is not the case, the cost of a cooperative robotic system to perform complex tasks, tends to remain less than a single complex robot.
\end{enumerate}

\subsection{Coordination Scheme of the Cooperative Robots}
In this section, we will discuss the three coordination schemes; master-slave control, centralized control, and decentralized control of the swarm robots. 

\begin{enumerate}
	\item \textbf{Master Slave Control: }In this scheme, one robot which is position control acts as a master and other are in compliant with it, means they are a slave to their master, to maintain the kinematic constraints. The advantage of this scheme is that each robot has its controller.
	\item \textbf{Centralized Control: }This scheme is based on centralized control, in which all the robots are connected, and they act as a closed kinematic chain. 
	\item \textbf{Decentralized Control: }This scheme has a decentralized control; all the robots work independently. It is better than master-slave configuration because it does not involve communication delay. They communicate with each other through sensors and are easy to implement than a centralized scheme.
\end{enumerate}

\begin{figure}
	\centering
	\includegraphics[width=1\linewidth]{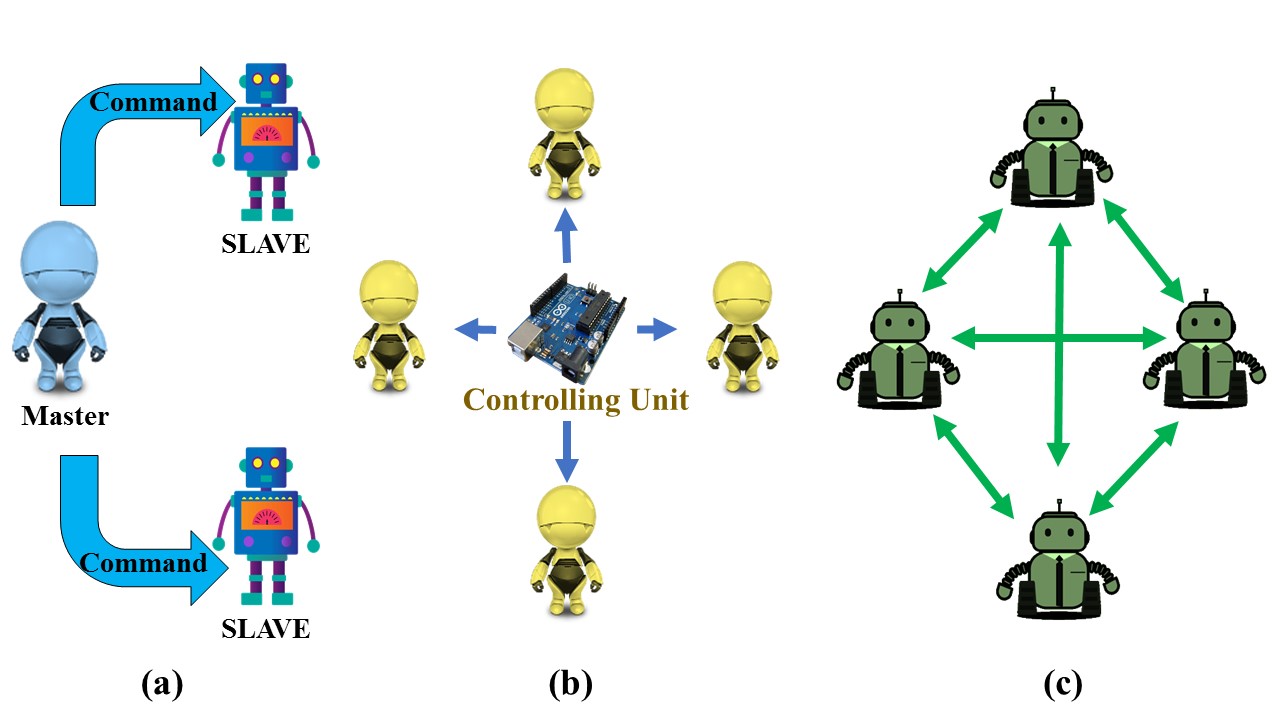}
	\caption{Figure (a), (b), and (c) are the master-slave, Centralized, and Decentralized Configurations of the swarm robots. (a) shows that in the army of a robot, there is a master robot that guides or leads the other robot and they are obedient to their master. (b) Shows that instead of a master, a centralized controlling unit is used to send all the signals to the robots. Finally, (c) tells us that robots work in a decentralized environment where everyone is master, without anyone's monopoly. \vspace{-0.5cm}}
	\label{Fig.5}
\end{figure}

All these three schemes are shown in Fig. \ref{Fig.5}, which shows that how they are different from each other and what are their pros and cons.

\subsection{Application of the Cooperative Robots}
In this sub-section, we will discuss the applications; rescue services, mining and agricultural services, espionage activities, underwear discoveries, war activities, and hiring services of the cooperative robots.
\begin{enumerate}
	\item \textbf{Rescue Missions: }Swarm robots are employed in rescue activities, during disastrous operations they are sent to the unreachable places to suspect the presence of life.
	\item \textbf{Mining and Agriculture: }They can be used to collect important data from mining areas, without endangering the life of a person. Likewise, they can collect data and the conditions of agricultural regions.
	\item \textbf{Espionage Purposes: }They are well suited for intelligence purposes. Swarm robots can be employed by military and intelligence agencies to collect the data from the terrorist or enemies’ territories \cite{35}.
	\item \textbf{Underwater Discoveries: }Cooperative robots can be used for the underwater discoveries, where a person can not reach, they can tell us about the atmospheric activities at a depth of the sea.
	\item \textbf{War Activities: }In the coming future, instead of using human resources the governments will deploy swarm robots in war-field. They can carry warheads and can work strategically to attack the targeted area.
	\item \textbf{Hiring Services: }The swarm robots can be used to acquire rental services. They can help in gathering information, data or other services to its acquirer.\par
	
\end{enumerate}
Some other worth mentioning practical applications from our past research work includes; distributed task allocation of multiple robots: A control perspective \cite{52}, decentralized kinematic control of a class of collaborative redundant manipulators via recurrent neural networks \cite{61}, cooperative distributed source seeking by multiple robots \cite{62}, Formation control and tracking for co-operative robots with non-holonomic constraints \cite{63},  multi-robot cooperative control for monitoring and tracking dynamic plumes \cite{64},  distributed recurrent neural networks for cooperative control of manipulator \cite{65}, decentralized control of collaborative redundant manipulators with partial command coverage via locally connected recurrent neural networks \cite{66}, distributed source seeking by cooperative robots: All-to-all and limited communications \cite{67}, cooperative motion generation in a distributed network of redundant robot manipulators with noises \cite{68}, and neural dynamics for cooperative control of redundant robot manipulators \cite{69}. 

\begin{figure}[tbh]
	\centering
	\includegraphics[width=1\linewidth]{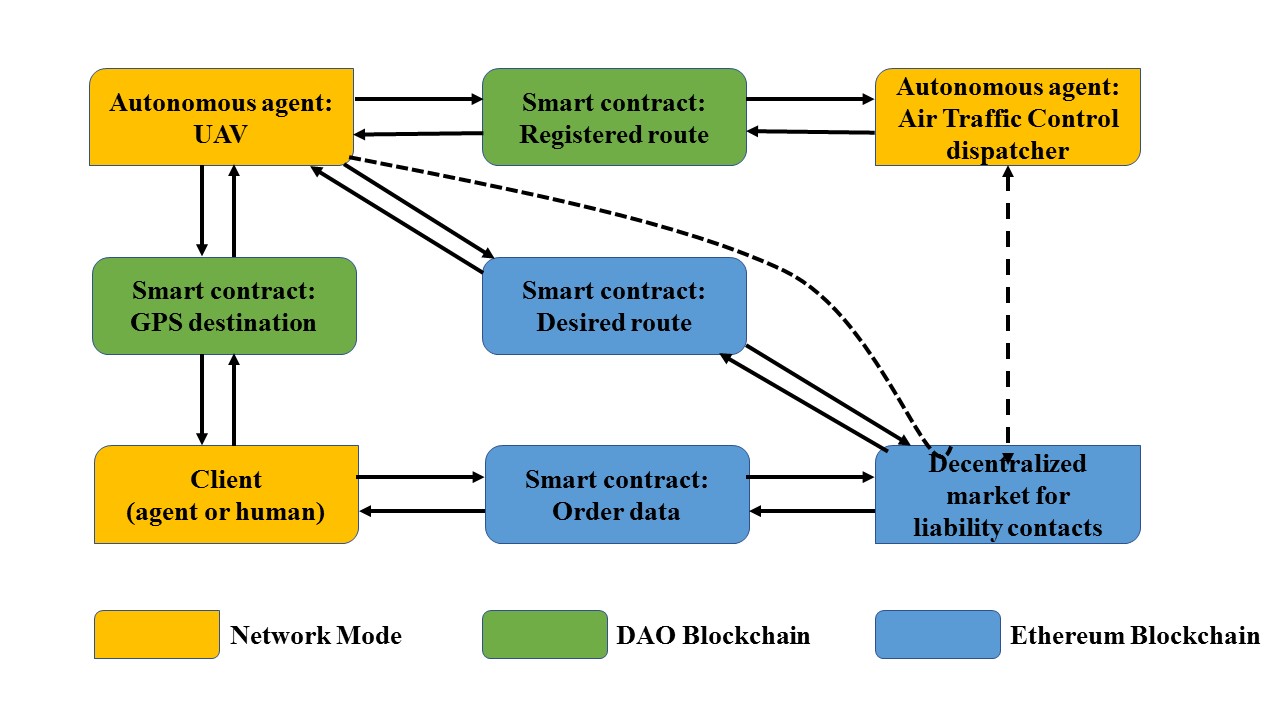}
	\caption{\textcolor{red}{This explains the architecture UAC based system. The dashed arrows shows the waiting for the contract to appear. All the blocks of network are integrated in decentralized environment.}}
	\label{Exp2}
\end{figure}
\subsection{Issues with Cooperative Robots}
In this section, we will discuss the issues like; security issues, distributed decision making, and variation in behavior, that researchers face in the field of cooperative robots.
\begin{enumerate}
	\item \textbf{Security Issues: }One of the main problem that swarm robots can face is the security issue since we have discussed their applications that require a lot of security and confidentiality of the data. In case of loose security, the data could be at stake. Safety is an important issue, because hacker or an intruder can virus one of the robots that can affect the whole army of robots, and can bring down the mission.
	\item \textbf{Distributed Decision Making: }In an army of swarm robots, distributed decision making plays a vital role. Each robot has different sensors to detect its surrounding, greater the number of robots more will be the viewpoints of the army. This will create an ambiguity, so an intelligent decision-making algorithm is required to fill this void.
	Researchers have done an enormous amount of work to fill this blank, which includes, dynamic tasks allocation, obstacle avoidance, and map building tasks. However, when the number of robots in a clan increases this task becomes difficult, so it's still an open problem to be addressed.
	\item \textbf{Variation in Behavior: }Another important aspect of the cooperative robotic system is the variation in their behavior like aggression, defensive, foraging, and flocking, {\it etc.} There should be a controlled-intelligent algorithm that can monitor their mood swings. 
\end{enumerate}

\subsection{Blockchain to resolve the Issues}
The blockchain with its decentralized and secure nature can facilitate the cooperative robots with their problems.
The solutions to the issues of cooperative robots with blockchain are listed below:
\begin{enumerate}
	\item \textbf{Security with Blockchain: }Researchers have done vigorous research on how to secure the communication between the army of the swarm robots. It is necessary because of the diverse nature of the system like; autonomy, decentralized control, and distributed control. The blockchain is a kind of technology that operates on peer-to-peer and decentralized connections, which makes their communication; safe, reliable, secure and immutable. Decentralization will make the swarm robots secure since it will make them independent of third-party intervention. Before, as we mentioned about the hashing and blockchain in \ref{s2} this will make the data communication secure because if someone tries to invade the privacy or alter the data, the hash of the block will change, which results in the breaking of a chain and makes the request invalid. Thus, causes the system free from potential threats, and vulnerabilities.
	\item \textbf{Distributed Decision Making and Blockchain: }Blockchain can help to tackle this problem, consider that the army of swarm robots are on a mission, and they are stuck in a situation where they need to make a decision. Let’s say one of the robots came up with the possible solution; blockchain will allow him to publish those possibilities in a block with a particular address to each option. The remaining robots will vote the solutions by accessing their addresses, the solution with maximum votes will come in action.
	\item \textbf{Behavior differentiation and Blockchain: }Blockchain can facilitate the swarm robots in this field as well. It provides the possibility to join different behavioral blocks in a hierarchy; this will help the robots to switch between different behavioral moods depending on their behavioral parameters like diversity, environment, permission, mining, {\it etc.} This technique is known as, pegged sidechain. We can replicate this to the other robots in a swarm as well.
	
\end{enumerate}
\begin{table*}[t]
	\begin{center}
		\caption{\textcolor{red}{This table comprehensively summarize the features, potential applications, limitations, and how bloackchain overcome those limitations of DCS and cooperative robots. It also points out the limitations of blockchain itself.}} \label{Table.1}
		\begin{tabular}{|l| l |l |l| l||l|}
			\hline
			Blockchain Applications & Features & Applications & Limitations & \shortstack{\\ Blockchain Overcome \\ Limitations} & Blockchain's Limitations\\
			\hline
			\hline
			& Handle complexities & Fluid-level control & Interception & Decentralization & Human Error \\
			Distributed control system& Pre-defined functions & Temperature control & Interruption  & Decentralization & Security lapses\\
			& Scalability & Pressure control & Modification & Ir-reversibility & Network size \\
			& Sophisticated HMI & PRDS control & Fabrication & Ir-reversibility & Latency\\
			\cline{1-5}
			& Distribution & Rescue operations & Security issues & Decentralization & Privacy \\
			& Decentralization & Mining/ Agriculture & Confidentiality & Peer-to-peer  & Encryption \\
			Cooperative robots& Robustness & Underwater discoveries & Decision making & Computational Logic & \\
			& Scalability & Military  & Behavior variation & Pegged sidechain & \\
			& Autonomous & Hiring services & Interception & Decentralization & \\
			\hline
		\end{tabular}
	\end{center}
\end{table*}
\textcolor{red}{
\subsection{Experimental results}
In this section, we will discuss the experimental results obtained, showing how blockchain can help the swarm robots in making decisions, coordination, and security. We will discuss two experiments; blockchain protocol for multi-agent systems includes UAVs \cite{79} and managing byzantine robots using blockchain \cite{78}.
\subsubsection{blockchain protocol for multi-agent systems includes UAVs}
Aleksandr et al. \cite{79} experimented, the idea was to make the communication between the multi-agent system more reliable and foolproof so that they can make decisions collectively in a more effective manner. Decentralization of the blockchain was the driving force between their experiment which ensures the safe and reliable communication between the multi-agent systems.\\
First, they developed the drone employee project using AIRA protocol, where UAVs are the multi-agent systems and to guide them while flight dispatchers were used. They established the communication between the UAVs and dispatchers that reserves the route points for them. They worked with topographic data and traced the route of UAV in the Digital Elevation Model(DEM) and stored it in the blockchain. UAV received their route point from the blockchain and made the flight.\\
They did the modification, and they divided the node system into two, Air Traffic Control (ATC) they perform the function of dispatchers and direct the UAVs for safe flight, and the second node is of Unmanned Traffic Management Balancer Watcher, it's a supervisor node to regulate the working of ATC nodes. The architecture of the system is shown in Fig. \ref{Exp2}. The experiment was that one of the users made a delivery request to the blockchain, available UAV accepted the request from blockchain and the special token was assigned to the user and UAV. The token included the user data and its location; the UAV made that information available to the dispatchers. Available dispatcher agreed with the UAV which included the location of the client, UAV followed the path provided by the dispatcher, and on its return, it informed the dispatcher about the completion of the mission. This experiment was concluded successfully.}

\textcolor{red}{
\subsubsection{Managing byzantine robots using blockchain}
The experiment was conducted by Volker et al. \cite{78}; the goal was to spread-out the robots over the provided surface, the surface contained two-color square boxes; black, and white. Robots task was to determine the most frequent color available in the environment.\\
For this experiment and keep track of the robot's behavior using low-level routines, the blockchain keeps the identity of all the robots, each robot keeps a copy of the blockchain and acts like a node and a miner in the blockchain. The decision to identify the frequency of color was based on consensus; blockchain provided them the space to share knowledge, record vote, and make a collective decision.\\}

\section{Limitations of Blockchain}\label{s5}
Although blockchain has paved its way into the technological and business world and we have even discussed its practical implementation in the field of distributed systems and cooperative or swarm robots.
Despite its practical implementation it still has some drawbacks that limit its implication. Some of its limits are discussed below:

\subsection{Human Error}
If the blockchain is used for storage, then data stored should be of 100\% trust-worthy, and of high-quality. Blockchain does not accommodate or accept the low-quality or false data. If you try to enter a false data, it will automatically discard the data. The phrase “garbage in garbage out,”truth about the blockchain, because it does not hold the falsified data.

\subsection{Unavoidable Security Flaw}
There is one unavoidable flaw in bitcoin and other blockchain technology, if the majority of the peers connected to the network tell a lie, then that lie becomes truth. Let’s say, 100 peers are connected to the network, and 51 of those somehow attack the system and tries to alter the data, if they do so, they can breach the security of the blockchain easily. This is called unavoidable or 51\% attack. That’s why blockchain communities observe the pools very carefully so that no can take over the pool quickly. Another way to avoid this attack is, the large pool, means that the number of peers working in a pool should be more in a name. For example, in the case of bitcoin, pool with 5000 peers is considered as a safe pool.

\subsection{Network Size}
Blockchains are very much welcome to the bad actors; they are friendly to them, that’s why they are called, “anti-fragile” system.
In our previous problem, we suggested that the pool should be ample to avoid “51\% attack”, but in doing so, peers need to take care of it that network should be robust with a widely distributed grid of nodes.

\subsection{Latency}
In \ref{s2}, we talked about the admittance of block to the blockchain, since it is time-dependent so after some time the block becomes the permanent member of the chain, and data stored in it becomes unalterable, immutable, or immune to any change. This cause the latency in the transaction of data. For example, in the case of bitcoin it is 10 min. It means that if someone did a transaction, it would become a permanent member of the chain after 10 mins, which is not likable or ideal. In the case of Ethereum, this time is less, only 10 to 15 sec. 

\subsection{Privacy}
Privacy of personal data is also one of the limitations of the blockchain. The openness of the data is not preferable all the time, who wants to share their bank statements with others? No one, so this is one of the flaw or security issues in the blockchain, it is also bad for those who misuse the public funds for their benefits.

\subsection{Encryption}
Encryption is useful to secure the data, in blockchain if the public key is made public everyone can use data to see the data. In case, if someone lost the key, he won’t be able to access his wallet ever in life. New technologies are underway to tackle this security hurdle of encryption, like; quantum computers are making their way in the technological world, and they will be able to crack any encrypted key in no time. This is a topic of discussion that how they can retain the prestige of encrypted data.

\section{Conclusion}
In this paper, we concluded that, despite the disruptive technology of swarm robots and DCS, they still face privacy and security challenges. To overcome these challenges blockchain facilitate them with its decentralization, peer to peer network configuration. Its decentralization removes the monopoly of the third party and creates a secure space in an un-trustful environment, which makes it ideal. Although these technological and innovative changes are in the making in future, blockchain can revolutionize the field of DCS systems and swarm robots and will provide them with a more secure and foolproof environment.

\balance

\newpage
%

\end{document}